\documentclass[aps,prl,twocolumn,superscriptaddress
 amsmath,amssymb,
 aps,
]{revtex4-2}

\usepackage{graphicx}% Include figure files
\usepackage{dcolumn}% Align table columns on decimal point
\usepackage{bm}% bold math
\usepackage{hyperref}% add hypertext capabilities

\usepackage{color}
\usepackage{subfigure}

\begin{document}

\title{Shear-Induced Electrophoretic Migration Perpendicular to the Electric Field}

\author{Andrés Rodríguez-Galán, Raúl Fernández-Mateo, Pablo García-Sánchez, Antonio Ramos.}

\affiliation{Depto. Electrónica y Electromagnetismo, Universidad de Sevilla, Seville, Spain.}

%\email{}


\begin{abstract}
Recent experiments combining electrophoresis with pressure-driven flows in microchannels have revealed that microparticles undergo lateral migration perpendicular to the applied electric field. Although fluid inertia has been proposed as a possible explanation, inertial effects are negligibly small in these regimes, leaving the underlying physical mechanism an open question. In this study, we address these observations by extending previous theoretical work on concentration polarization,i.e., the external-field-induced modification of the ionic concentration field surrounding a dielectric object. We consider a dielectric particle with surface conductance subjected simultaneously to an external electric field and a shear flow. We show that the shear flow breaks the symmetry of the ionic concentration around the particle in the direction perpendicular to the applied field, thereby driving lateral migration. We demonstrate that the resulting migration velocity comprises two distinct contributions: an electrophoretic and a diffusiophoretic component. Our theory yields an explicit expression for the velocity magnitude as a function of the zeta potential and the Dukhin number, predicting typical speeds on the order of $\mathrm{\mu}$m/s for representative experimental parameters. Notably, the model also predicts a reversal in the migration direction for Dukhin numbers of order unity.
\end{abstract}

\maketitle
% \begin{keywords}
% Electrokinetics, Electrophoresis, Cross-Stream Migration, Concentration Polarization, Particle Focusing, Microfluidics.
% \end{keywords}

%\section{Introduction}
%\label{intro}

Electrophoresis describes the motion of a charged solid particle suspended in an electrolyte under the action of an externally applied electric field \cite{hunter}. The electrophoretic velocity is typically characterized through the Helmholtz--Smoluchowski relation,
\[
\bm{U}_{\mathrm{ep}} = \frac{\varepsilon \, \zeta_0}{\eta}\,\bm{E},
\]
where \(\varepsilon\) denotes the permittivity of the electrolyte, \(\zeta_0\) is the particle zeta potential, \(\eta\) represents the dynamic viscosity of the fluid, and \(\bm{E}\) is the imposed electric field \cite{smoluchowski}. This formulation applies to moderately charged particles when the electrical double layer (EDL) is thin compared with the characteristic particle size. The EDL comprises the surface charge located at the particle interface together with the surrounding diffuse ionic cloud that electrostatically screens this charge in the electrolyte \cite{hunter}.

Electrophoresis has been extensively studied and applied across a wide range of fields, including gel electrophoresis for biomolecular separation \cite{zhu2012} and microfluidic transport and flow control \cite{Stone04}. More recently, experimental studies have explored the combined action of electric fields and pressure-driven flows in microfluidic channels \cite{vishwanathan2023, Abdorahimzadeh2023, Abdorahimzadeh2024,li2018}. These studies revealed that micro- and nanoparticles can migrate in a direction perpendicular to both the electric field and the mean flow direction. Depending on the operating conditions, particles migrate either toward the channel centerline or toward the channel walls, highlighting the potential of this mechanism for particle focusing, enrichment, and fractionation.

Besides its potential applications in the manipulation and separation of small particles, the physical mechanism underlying this cross-stream migration remains poorly understood. Lateral migration in microchannels is commonly associated with inertial effects \cite{DiCarlo2007}, which are widely exploited in microfluidic particle fractionation. However, we emphasize that the migration discussed here does not occur under pressure-driven flow alone, but emerges only when the flow is combined with an applied electric field. Recent works \cite{khair2020,Choudhary2019} have investigated the role of inertial effects on a particle suspended in a flow generated by the superposition of shear and electrophoresis. Khair and Kabarowski \cite{khair2020} derived an expression for the lateral migration velocity, predicting values that are exceedingly small under typical experimental conditions.

These findings suggest that additional physical mechanisms must be responsible for the particle migration perpendicular to both the electric field and the mean flow direction, occurring only when the two fields act simultaneously.
\begin{figure}[t!]
    \centering
    \includegraphics[width=1\linewidth]{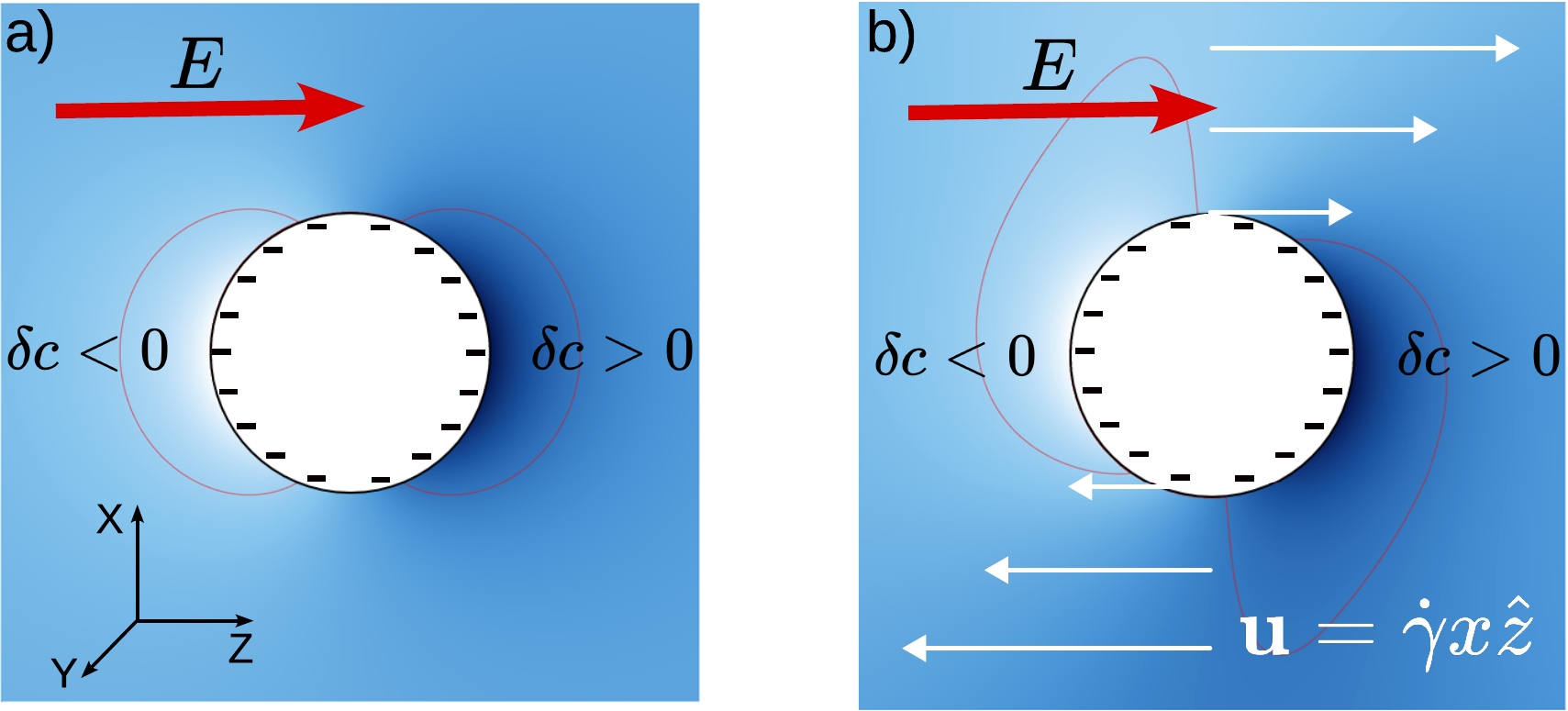}

\caption{a) The colormap shows the concentration polarization around a spherical particle suspended in a quiescent electrolyte. The ionic concentration increases on one side of the particle ($\delta c > 0$) and decreases on the opposite side ($\delta c < 0$).
b) Concentration polarization around a spherical particle subjected to a linear shear flow. Advection of ions by the flow perturbs the concentration field, thereby breaking the symmetry with respect to the YZ plane.}
    \label{figure1}
\end{figure}
In this work, we investigate the phenomenon of concentration polarization (CP) around a colloidal particle subjected to an external electric field \cite{dukhinshilovbook}. CP refers to the development of ionic concentration gradients in the electrolyte surrounding the particle. These gradients originate from the particle surface conductance: when an electric field is applied, the electromigration of counterions within the electrical double layer (EDL) induces ion depletion on one side of the particle and ion enrichment on the opposite side. Figure \ref{figure1}a) schematically illustrates CP around a negatively charged spherical particle suspended in an electrolyte. The ionic concentration field shown in Fig.~\ref{figure1}a) results from the balance between electromigration currents along the particle surface and diffusive fluxes driven by the resulting concentration gradients.

We are interested in modeling a particle subjected simultaneously to an electric field and a flow field. To this end, we consider a linear shear flow of the form  $\textbf{u}=\dot{\gamma}x\hat z$, where $\dot{\gamma}$ denotes the shear rate. In pressure-driven microchannel flows, the velocity profile is generally Poiseuille-like rather than purely linear. However, when the particle size is much smaller than the channel dimensions, the local flow field around the particle can be reasonably approximated by a linear shear flow. As will be shown in this work, the fluid velocity enters the convection--diffusion equation governing the ionic concentration field. As a consequence of ion advection, the concentration distribution loses its symmetry with respect to the YZ plane. This effect is illustrated schematically in Fig.~\ref{figure1}b), and its implications will be explored throughout the remainder of the paper. We anticipate here that this symmetry breaking in the ionic concentration field gives rise to an electrophoretic velocity that is not strictly aligned with the applied electric field. In addition to the conventional electrophoretic motion parallel to the field, the particle acquires a velocity component perpendicular to the electric field direction.\\

\textbf{Theory}. We consider a dielectric sphere bearing a negative surface charge density $q_s$ and immersed in a binary symmetric electrolyte (i.e. positive and negative ions have the same diffusivities, $D_+=D_-=D_0$). The system is subjected to a DC electric field and a linear shear flow and our goal is to calculate the electrophoretic velocity of the sphere. We adopt the framework developed by Schnitzer and Yariv for electrokinetic flows at large zeta potentials \cite{Schnitzer2012} (hereafter referred to as the SY model). This approach has been successfully applied to describe particle electrophoresis \cite{Schnitzer2013PhysFluids,Schnitzer2014PhysFluids}, as well as quadrupolar flows around pillars \cite{Calero2021PRAppl} and spheres subjected to AC fields \cite{fernandez2021}. The SY model establishes the governing equations and boundary conditions for the electric potential, $\phi$, ionic concentration, $c$, and fluid velocity, $\textbf{u}$, in the bulk electrolyte assuming that the EDL is thin compared with the characteristic particle size. This assumption is valid for the experiments mentioned above where microscopic particles are dispersed in electrolytes with typical values of the Debye length in the range 10-100 nm.

We adopt a reference frame fixed at the center of mass of the particle, and express the governing equations in nondimensional form using the following characteristic scales: the particle radius $a$ as the length scale; the bulk ionic concentration $c_0$ as the concentration scale; and the thermal voltage for the electric potential, $V_{\mathrm{th}}=k_{\mathrm {B}} T/{ze}$, where $T$ is the absolute temperature, $k_{\mathrm B}$ is the Boltzmann constant, $z$ is the ionic valence, and $e$ is the elementary charge. For a symmetric $1{:}1$ electrolyte at room temperature, the thermal voltage is $V_{\mathrm{th}}\approx 25\,\mathrm{mV}$. Velocities are scaled by $\varepsilon V_{\mathrm{th}}^2/(a\eta)$, which in turn defines the characteristic time scale as $a^2\eta/(\varepsilon V_{\mathrm{th}}^2)$. The non-dimensional governing equations are:
\begin{eqnarray}
D\nabla^2c-\textbf{u}\cdot\nabla c&=&0,\label{condif}\\
\nabla\cdot(c\nabla\phi)&=&0, \label{corriente}\\
-\nabla p+ \nabla^2{\mathbf{u}}+\nabla^2\phi\nabla\phi&=&0, \label{stokes1}\\
\nabla\cdot\mathbf{u}&=&0. \label{stokes2}
\end{eqnarray}
Eq. (\ref{condif}) is the convection--diffusion equation for the ion concentration with $D=D_0/(\varepsilon V^2_{\mathrm{th}}/\eta)$, while the equation for the electric potential (\ref{corriente}) corresponds to the continuity of the electrical current. Eqs. (\ref{stokes1}) and (\ref{stokes2}) are the Stokes equations for the fluid velocity (negligible Reynolds number) including a body force term of electrical origin.\\

The SY model also establishes the boundary conditions at the particle surface. The model assumes that the particle zeta potential is much larger than the thermal voltage, i.e., $\zeta_0 \gg 25\,\mathrm{mV}$, an assumption that is well satisfied in the experiments described above involving carboxylate microspheres. Under these conditions, co-ions are effectively excluded from the diffuse layer, while the counter-ion flux normal to the surface is balanced by the surface current within the EDL. As a result, the following boundary conditions are obtained:
\begin{eqnarray}\label{bcs}
\left(\frac{\partial c}{\partial n}- c\frac{\partial \phi}{\partial n}\right) &=& 0,\quad \mathrm{at}\, r=1, \\
\left(\frac{\partial c}{\partial n}+ c\frac{\partial \phi}{\partial n}\right) &=& -2\mathrm{Du}\nabla_s^2\left(\phi+\ln c\right), \quad \mathrm{at}\, r=1,
\end{eqnarray}
where $\nabla_s$ denotes the surface gradient operator tangent to the particle, and $\mathrm{Du}$ is the dimensionless Dukhin number, which quantifies the relative contribution of surface to bulk currents \cite{delgado05}.

Concerning the fluid velocity at the particle surface, the SY model prescribes the following slip boundary condition, which coincides with the expression derived in \cite{Prieve1984JFM} in the context of diffusiophoresis:
\begin{eqnarray}
{\textbf{u}}^s&=&\zeta\nabla_s\phi - 4\ln\left[\cosh(\zeta/4)\right]\nabla_s \ln c,
\end{eqnarray}
where the non-dimensional zeta potential $\zeta$ is related to the non-dimensional surface charge density, $q_s$, and ion concentration through the Gouy-Chapman equation $q_s=2\sqrt{c}\sinh{(\zeta/2)}$, \cite{hunter}. 

The boundary condition for the fluid velocity at the particle surface is:
\begin{eqnarray}
{\textbf{u}}={\textbf{u}}^s+\mathbf{\Omega}\times\textbf{r}, \quad \mathrm{at}\quad r=1
\end{eqnarray}
where $\mathbf{\Omega}$ is the angular velocity of the spherical particle.

In the far-field limit ($\mathbf{r}\rightarrow\infty$), the electric potential approaches that of the imposed field, $\phi = -\beta z$, where $\beta = E_0(a/V_{\mathrm{th}})$, while the ionic concentration relaxes to its bulk value, $c = 1$. Since the reference frame is attached to the particle center of mass, the far-field fluid velocity is given by the superposition of the imposed shear flow and the electrophoretic translational velocity $\mathbf{U}$, namely, $\mathbf{u}=\Gamma x \hat{z}-\mathbf{U}$, with $\Gamma=(\varepsilon V_{\mathrm{th}}^2 / a^2\eta)\dot{\gamma}$. The unknown translational and angular velocities, $\mathbf{U}$ and $\mathbf{\Omega}$, are determined by imposing vanishing net hydrodynamic force and torque on the particle. These conditions are expressed as $\oint_{S} \tilde{\mathbf{T}}\cdot d\mathbf{S} = 0$ and $\oint_{S} (\textbf{r}\times\tilde{\mathbf{T}})\cdot d\mathbf{S} = 0$, where $\tilde{\mathbf{T}}$ is the hydrodynamic stress tensor and the integrals are evaluated over a closed surface enclosing the sphere and located sufficiently far from it.

Note that the SY model was originally derived for a particle immersed in a quiescent fluid, so its applicability to situations involving an externally imposed flow may reasonably be questioned, since fluid motion could alter ion transport within the EDL and, consequently, modify the boundary conditions (\ref{bcs}). Nevertheless, following the analysis in \cite{yariv2011} for the streaming potential generated by a falling sphere, the contribution of charge advection scales as $q_s u \delta$, where $\delta$ is the non-dimensional Debye length. In the thin-EDL limit and typical experimental conditions, this contribution is negligibly small compared with the electromigration current associated with surface conductance. Therefore, the boundary conditions (\ref{bcs}) remain valid.\\

\textbf{Results.} We seek an analytical expression for the particle velocity in the limit of weak electric and shear forcing. To this end, the governing fields are expanded asymptotically in powers of the dimensionless electric-field strength, $\beta$, and the dimensionless shear rate, $\Gamma$:
\begin{eqnarray}
c&=&1+\beta c_{01}+\Gamma\beta c_{11}+\dots,\\
\phi&=&\beta \phi_{01}+\Gamma\beta \phi_{11}+\dots,\\
\mathbf{u}&=&\beta \mathbf{u}_{01}+\Gamma \mathbf{u}_{10}+\Gamma\beta \mathbf{u}_{11}+\dots,
\end{eqnarray}
where the first subscript denotes the order of the expansion in $\Gamma$, while the second denotes the order in $\beta$. The velocity field $\mathbf{u}_{01}$ corresponds to pure electrophoresis, i.e., the flow generated in the absence of an imposed shear flow, whereas $\mathbf{u}_{10}$ represents the flow around a sphere subjected to a linear shear. Neither of these contributions induces particle motion perpendicular to the electric-field direction. We therefore focus on the next term in the expansion, $\mathbf{u}_{11}$, which can, in principle, produce a non-zero lateral migration velocity proportional to the product $\Gamma\beta$.

In the Supplementary Material, we detail the derivation of the slip velocity at this order, $\mathbf{u}^s_{11}$. The calculation involves lengthy algebraic manipulations; notably, because the regular perturbation scheme fails for the concentration field \cite{leal2007}, the method of matched asymptotic expansions is required to determine $c_{11}$ and $\phi_{11}$. By invoking the Lorentz reciprocal theorem for Stokes flows \cite{masoud2019}, the velocity of the spherical particle can then be obtained as the surface average of the slip velocity. Since our interest lies in the $x$-component of the particle velocity, $U_x$,we write:
\begin{equation}
\frac{U_x}{\Gamma\beta}=-\left(\frac{1}{4\pi}\int_{S_p}\mathbf{u}^s_{11}\mathrm{d}S\right)\cdot\hat x
\end{equation}
where the integration is performed over the particle surface, $S_p$.

\begin{figure}[t]
    \centering
    \includegraphics[width=1\linewidth]{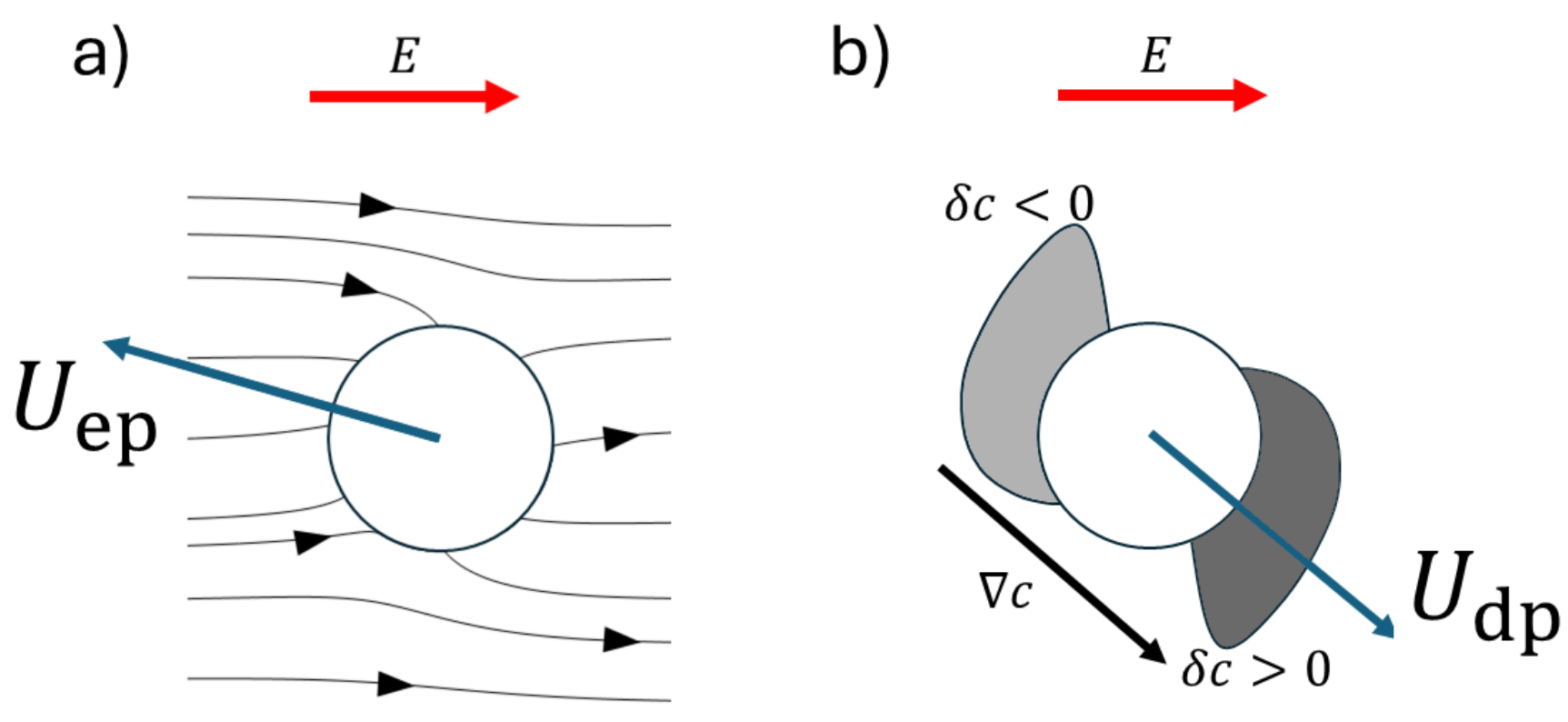}

\caption{\textbf{a)} Electrophoretic velocity, $\textbf{U}_{\mathrm{ep}}$, resulting from the combination of the applied electric field and the field induced by concentration polarization (as perturbed by the shear flow), $\textbf{E}=-\beta(\nabla\phi_{01}+\Gamma\nabla\phi_{11})$. \textbf{b)} Diffusiophoretic velocity, $\textbf{U}_{\mathrm{dp}}$, arising from the concentration polarization gradient, $\delta c=\beta(c_{01}+\Gamma c_{11})$. The illustrations correspond to $\Gamma$ values unrealistically high; but it clearly illustrates the underlying mechanisms.}
    \label{mecanismos}
\end{figure}

The non-dimensional expression for the cross-stream particle velocity is obtained as:
\begin{equation}
\small \frac{U_x}{\Gamma\beta/D}=\frac{\mathrm{Du}}{(1+2\mathrm{Du})^2}\left[\frac{|\zeta_0|\mathrm{Du}}{4}-(1+\mathrm{Du})\ln(\cosh(\zeta_0/4)) \right]
\end{equation}
where two distinct contributions to the velocity can be identified (see Fig. \ref{mecanismos}): the first term represents a purely electrophoretic contribution arising from the induced electric field along the $x$-direction, whereas the second stems from diffusiophoresis caused by the asymmetry in the induced ionic concentration gradients.\\

\begin{figure}[h]
    \centering
    % Primera subfigura
    \begin{subfigure}
        \centering
        \includegraphics[width=0.45\textwidth]{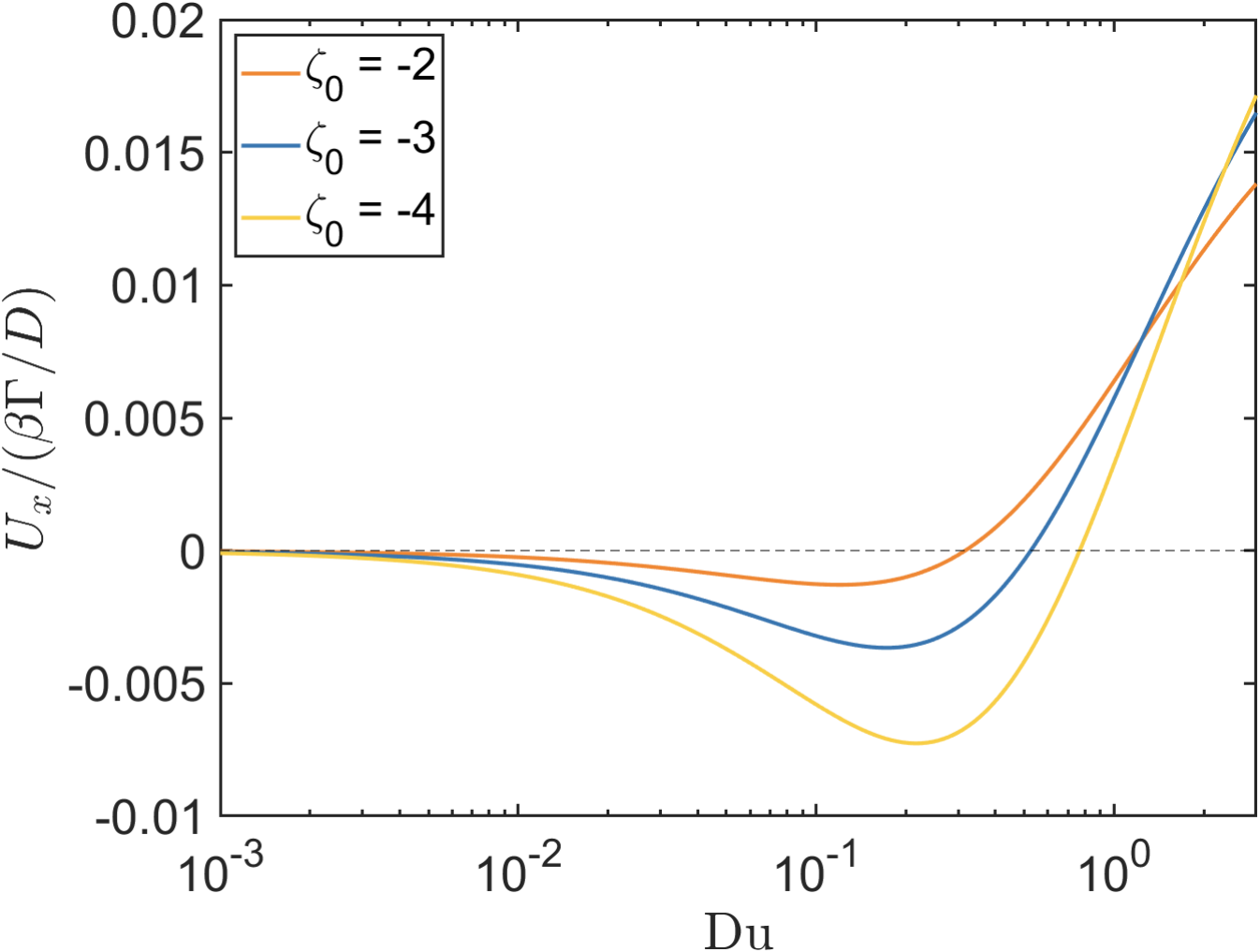}
        \caption{Dimensionless cross-stream migration velocity as a function of the Dukhin number for three representative values of $\zeta_0$. Positive values of $U_x$ indicate migration toward regions with a higher fluid velocity along the direction of the applied electric field. The velocity exhibits a change of sign as the Dukhin number increases.}
        \label{figure2}
    \end{subfigure}
    
    \vspace{0.5cm} % Añade un poco de espacio vertical de separación
    
    % Segunda subfigura
    \begin{subfigure}
        \centering
        \includegraphics[width=0.5\textwidth]{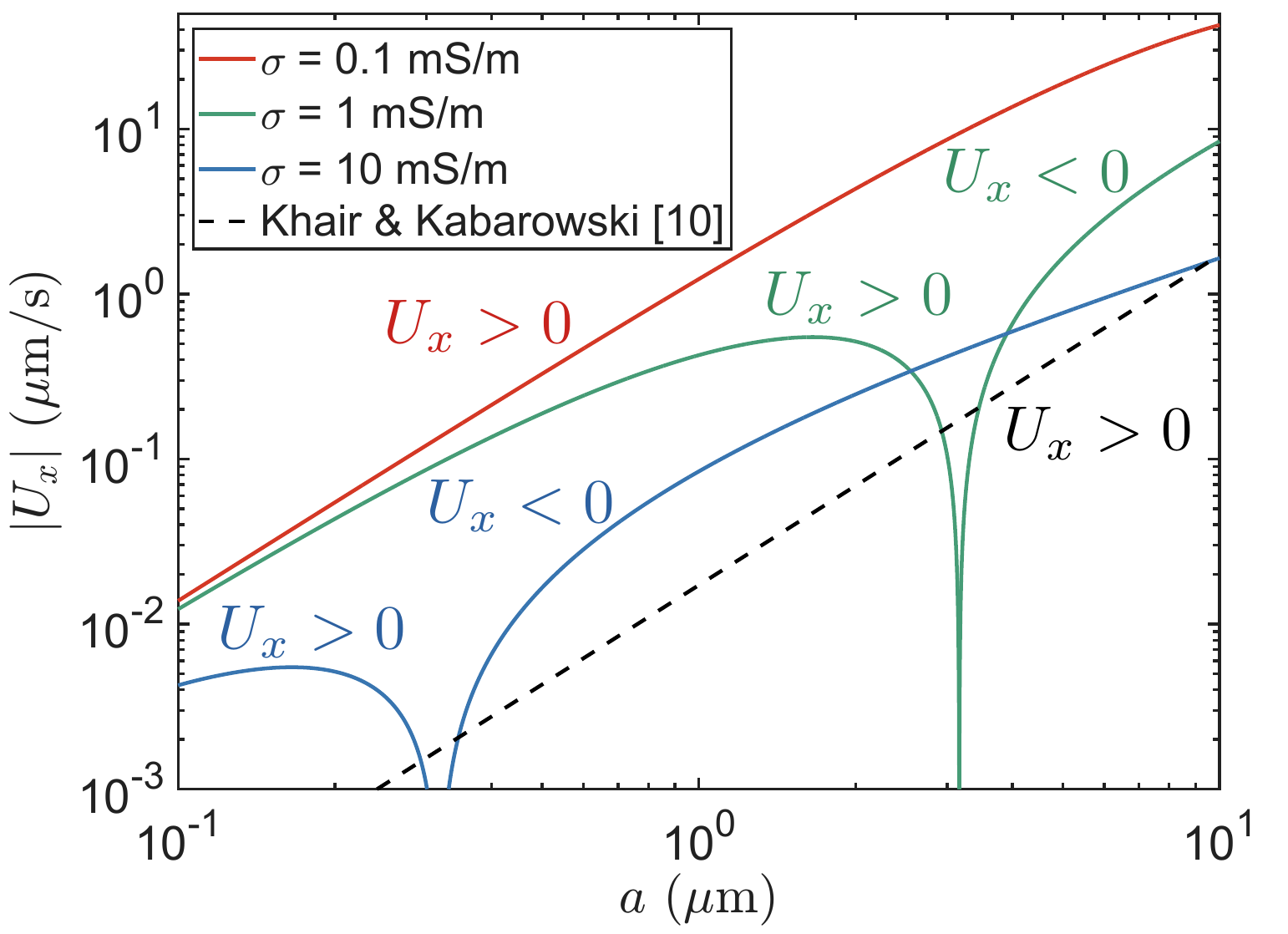}
        \caption{Cross-stream migration velocity as a function of particle radius for three electrolyte conductivities: $\sigma = 0.1$, $1$, and $10\ \mathrm{mS/m}$. The remaining physical parameters are $\zeta = -50\ \mathrm{mV}$, $D_0 = 2 \times 10^{-9}\ \mathrm{m^2/s}$, $E_0 = 30\ \mathrm{kV/m}$, $\dot{\gamma} = 55.6\ \mathrm{s^{-1}}$, and $K_s = 1\ \mathrm{nS}$. The theoretical prediction for the inertial electrophoretic lift by Khair and Kabarowski~\cite{khair2020} is also plotted for comparison.}
        \label{figcomparison}
    \end{subfigure}
   
\end{figure}

Figure~\ref{figure2} shows $U_x$ as a function of the Dukhin number, $\mathrm{Du}$, for three representative values of the dimensionless zeta potential, $\zeta_0 = -2, -3$ and $-4$. Positive values of $U_x$ indicate cross-stream migration toward regions with a higher fluid velocity along the direction of the applied electric field. In microchannel experiments, this implies that particles migrate toward the channel center when the electric field is aligned with the fluid flow. Remarkably, $U_x$ reverses its sign as the Dukhin number increases.\\

%Figure~\ref{figure2}b presents $U_x$ as a function of $\zeta_0$ for several values of the Dukhin number, $\mathrm{Du}=\{0.1,1,10\}$.

% \begin{figure}[h!]
%     \centering
%     \includegraphics[width=1\linewidth]{figure2.pdf}
% \caption{\textbf{a)} Cross-stream migration velocity as a function of the Dukhin number for two representative values of $\zeta_0$. The velocity exhibits a change of sign as the Dukhin number increases. \textbf{b)} Cross-stream migration velocity as a function of $\zeta_0$ for representative values of the Dukhin number.}
%     \label{figure2}
% \end{figure}

Figure~\ref{figcomparison} shows the cross-stream migration velocity (in $\mu\mathrm{m/s}$) as a function of the particle radius for three typical experimental conductivities: $\sigma = 0.1$, $1$, and $10\ \mathrm{mS/m}$. The physical parameters are chosen to match the Newtonian fluid experiments by Li et al.~\cite{li2018}, namely $E_0 = 30\ \mathrm{kV/m}$, $\dot{\gamma} = 55.6\ \mathrm{s^{-1}}$, and $D_0 = 2 \times 10^{-9}\ \mathrm{m^2/s}$ (corresponding to KCl in water). Here, the Dukhin number is estimated as $\mathrm{Du} = K_s / (a\sigma)$, where $K_s$ denotes the surface conductance, typically on the order of $1\ \mathrm{nS}$ for colloids~\cite{ermolina05}, and a representative zeta potential of $-50\ \mathrm{mV}$ is assumed. For comparison, we have also plotted the theoretical prediction by Khair and Kabarowski~\cite{khair2020} for the inertial electrophoretic lift. For particles with a radius of around $1\ \mu\mathrm{m}$, this inertial effect remains much smaller than the cross-stream migration velocity induced by concentration polarization.\\

% \begin{figure}
%     \centering
%     \includegraphics[width=\linewidth]{FigComparison.pdf}
%     \caption{$\zeta = -50$ mV, $E_0 = 30$ kV/m, $\dot\gamma=277.8$ s$^{-1}$, $K_s=1$ nS.}
%     \label{figcomparison}
% \end{figure}

\textbf{Conclusions.} We have employed the macroscale model derived by Schnitzer and Yariv for thin electrical double layers to investigate the effect of an imposed shear flow on the concentration polarization around a spherical particle. We have shown that, as a consequence of the combined action of electric and shear flow fields, the particle undergoes electrophoretic migration perpendicular to both driving fields. The model predicts velocities on the order of $\mu\mathrm{m/s}$ for particles with radii around $1\ \mu\mathrm{m}$, in good agreement with experimental observations in microchannels where particles migrate across the channel cross-section on timescales of a few seconds. Remarkably, the theory also predicts a reversal in the sign of the cross-stream migration velocity with increasing Dukhin number ($\mathrm{Du}$). Specifically, in microchannel experiments where the electric field and fluid flow are aligned, particles migrate toward the walls at low $\mathrm{Du}$, whereas migration occurs toward the center at higher values. This behavior is qualitatively consistent with the experimental observations reported by Abdorahimzadeh et al.~\cite{Abdorahimzadeh2024}, who observed this reversal in migration direction upon increasing the electrolyte conductivity.\\

Future work will focus on systematic experiments for a rigorous quantitative comparison with the theory. In this context, accounting for wall effects in microchannel geometries will be essential; notably, wall repulsion induced by concentration-polarization electroosmosis~\cite{fernandez2022} may play a significant role at high electric field strengths. In addition, future studies should address the behavior of nanocolloids, for which the thin electrical double layer approximation is no longer valid.\\

%\begin{acknowledgments}
\textbf{Acknowledgments.} The authors acknowledge the financial support from MICIU/AEI/10.13039/501100011033/FEDER, UE (Grant No. PID2022-138890NB-I00). A.R.G. acknowledges support from a predoctoral contract funded by Junta de Andalucía, Spain (Grant No. DGP\_PRED\_2024\_01162). R.F.M. acknowledges funding from the European Union’s Horizon Europe research and innovation programme under the Marie Skłodowska-Curie (Grant No. 101149570).
%\end{acknowledgments}

%

\clearpage
\begin{figure}[p]
    \centering
    \includegraphics[width=0.9\textwidth,page=1]{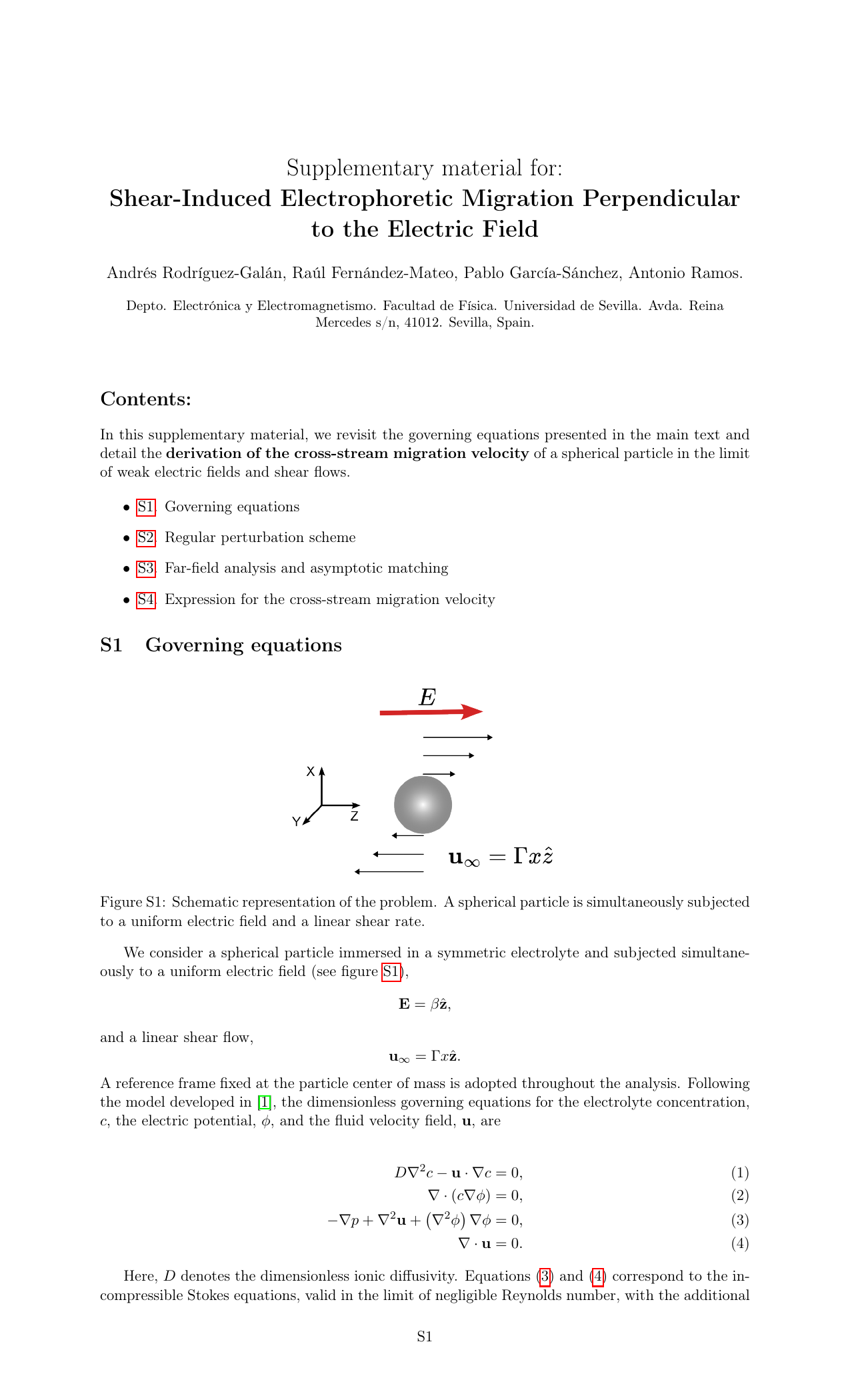}
\end{figure}

\clearpage
\begin{figure}[p]
    \centering
    \includegraphics[width=0.9\textwidth,page=2]{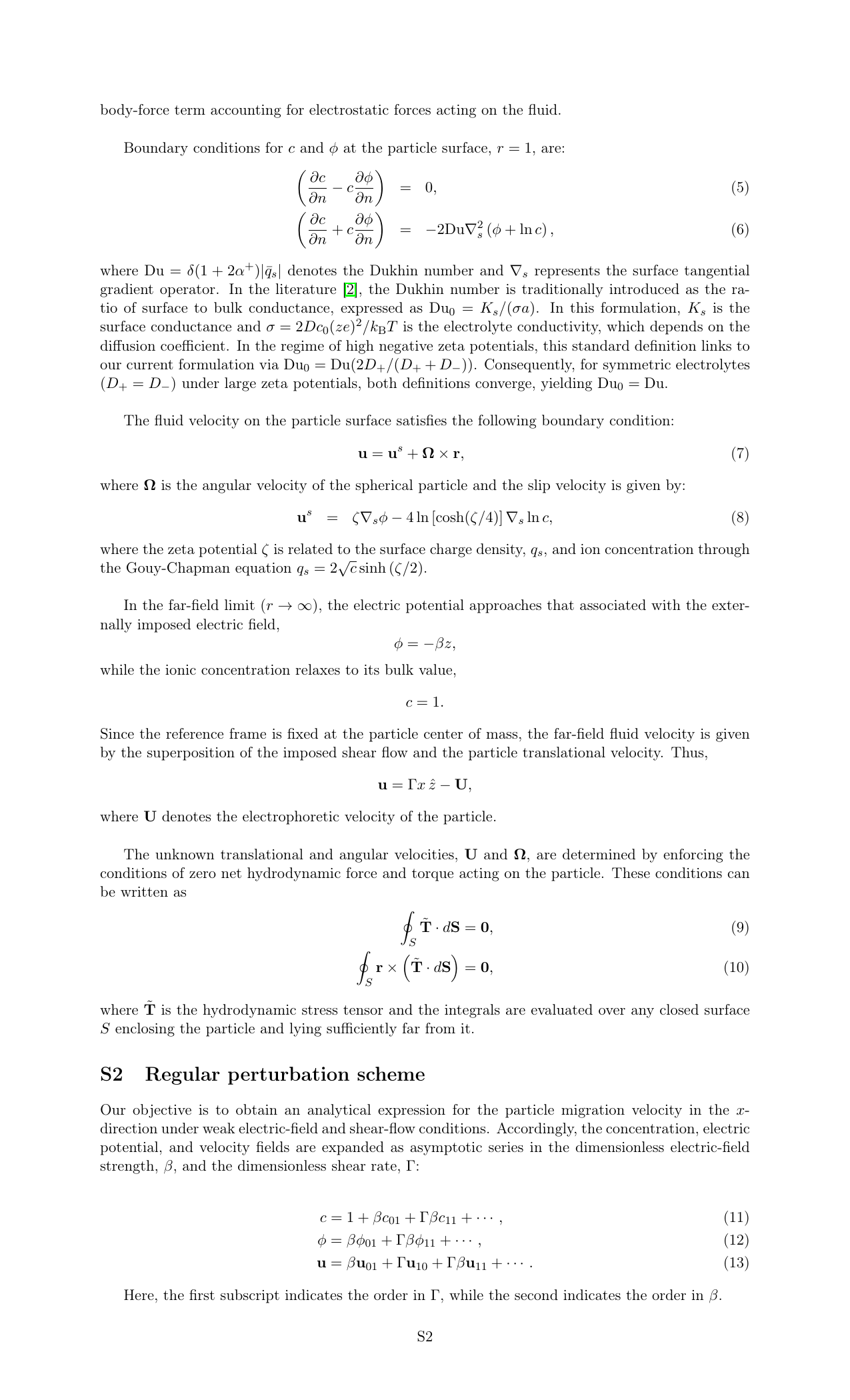}
\end{figure}

\clearpage
\begin{figure}[p]
    \centering
    \includegraphics[width=0.9\textwidth,page=3]{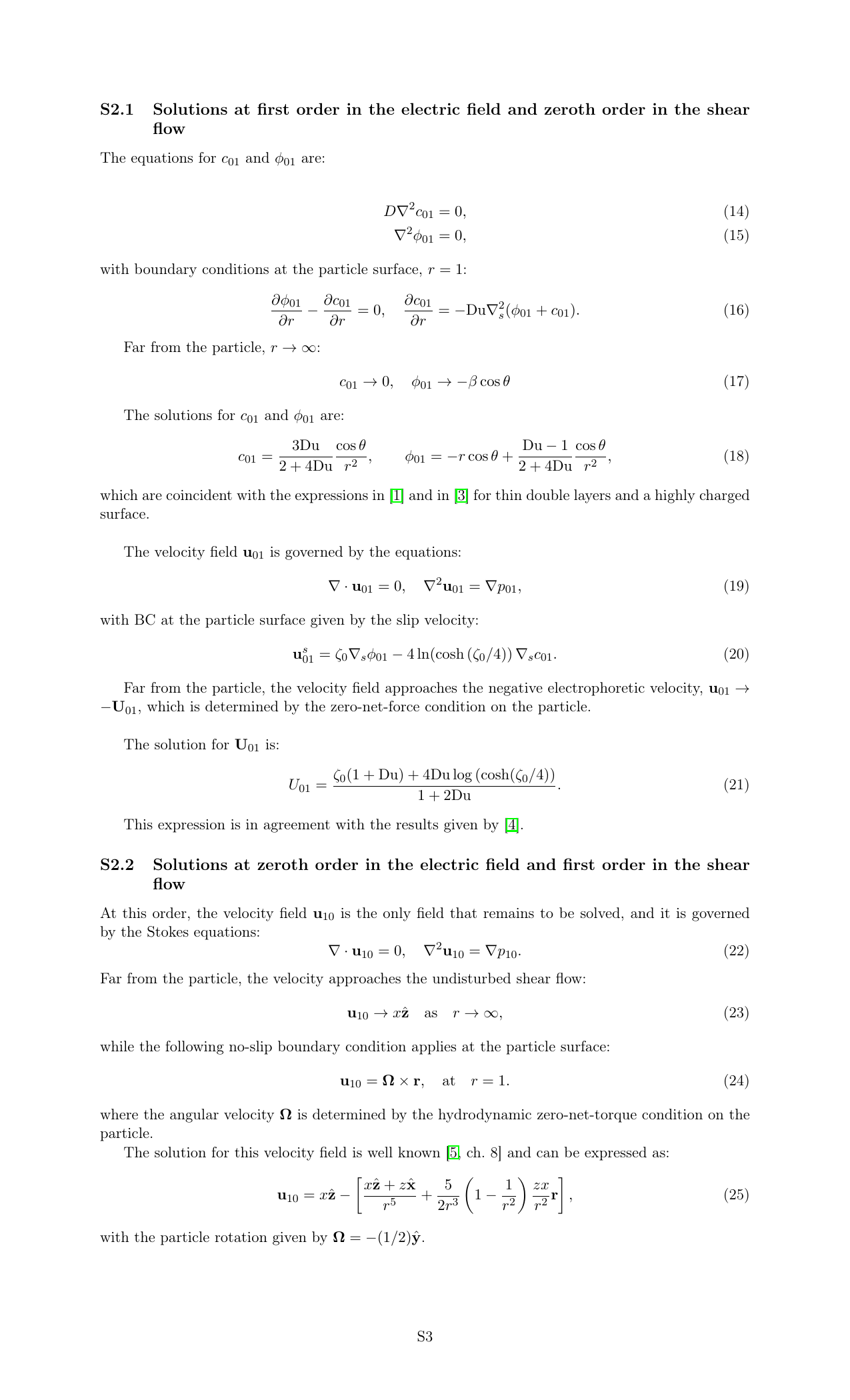}
\end{figure}

\clearpage
\begin{figure}[p]
    \centering
    \includegraphics[width=0.9\textwidth,page=4]{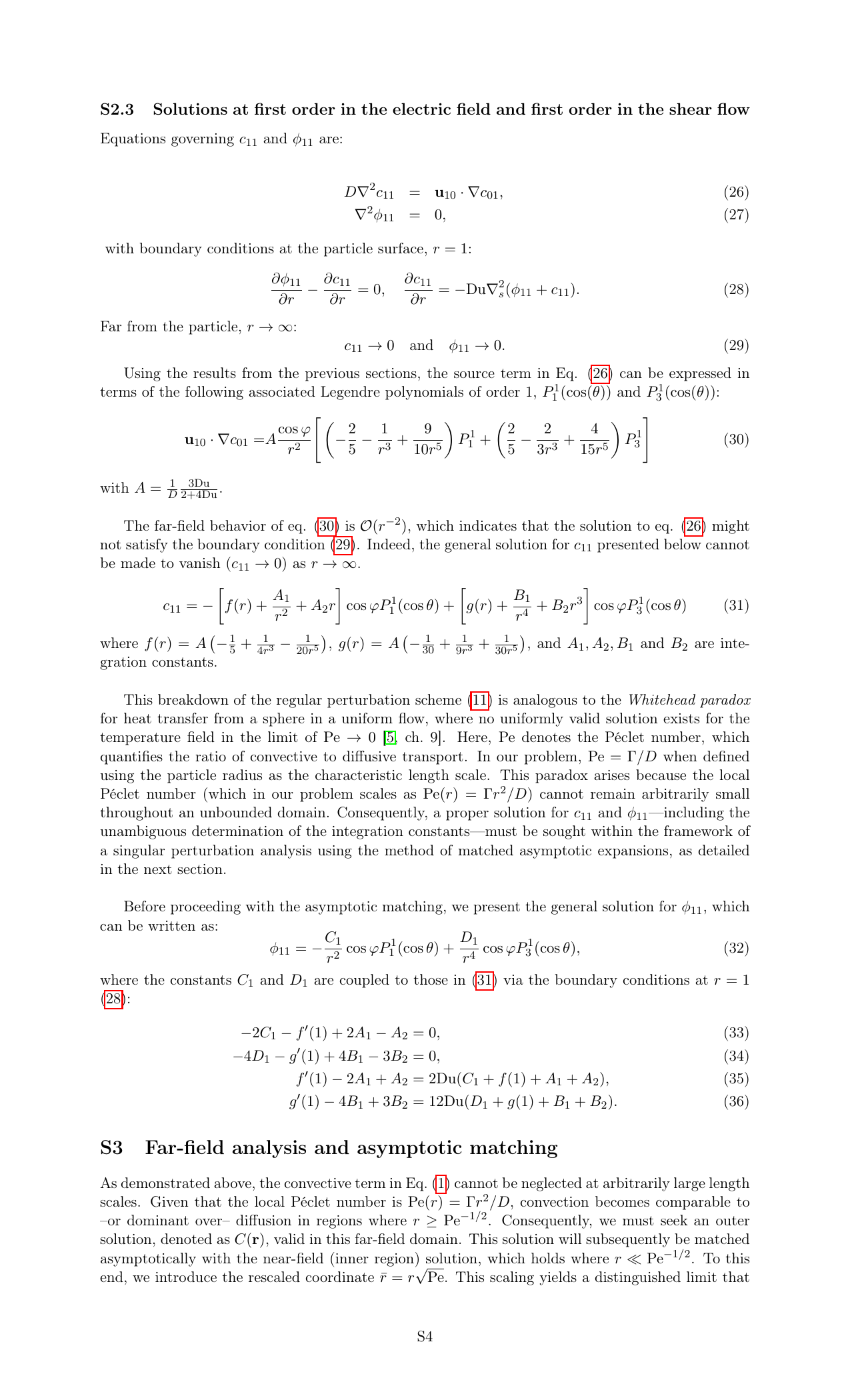}
\end{figure}

\clearpage
\begin{figure}[p]
    \centering
    \includegraphics[width=0.9\textwidth,page=5]{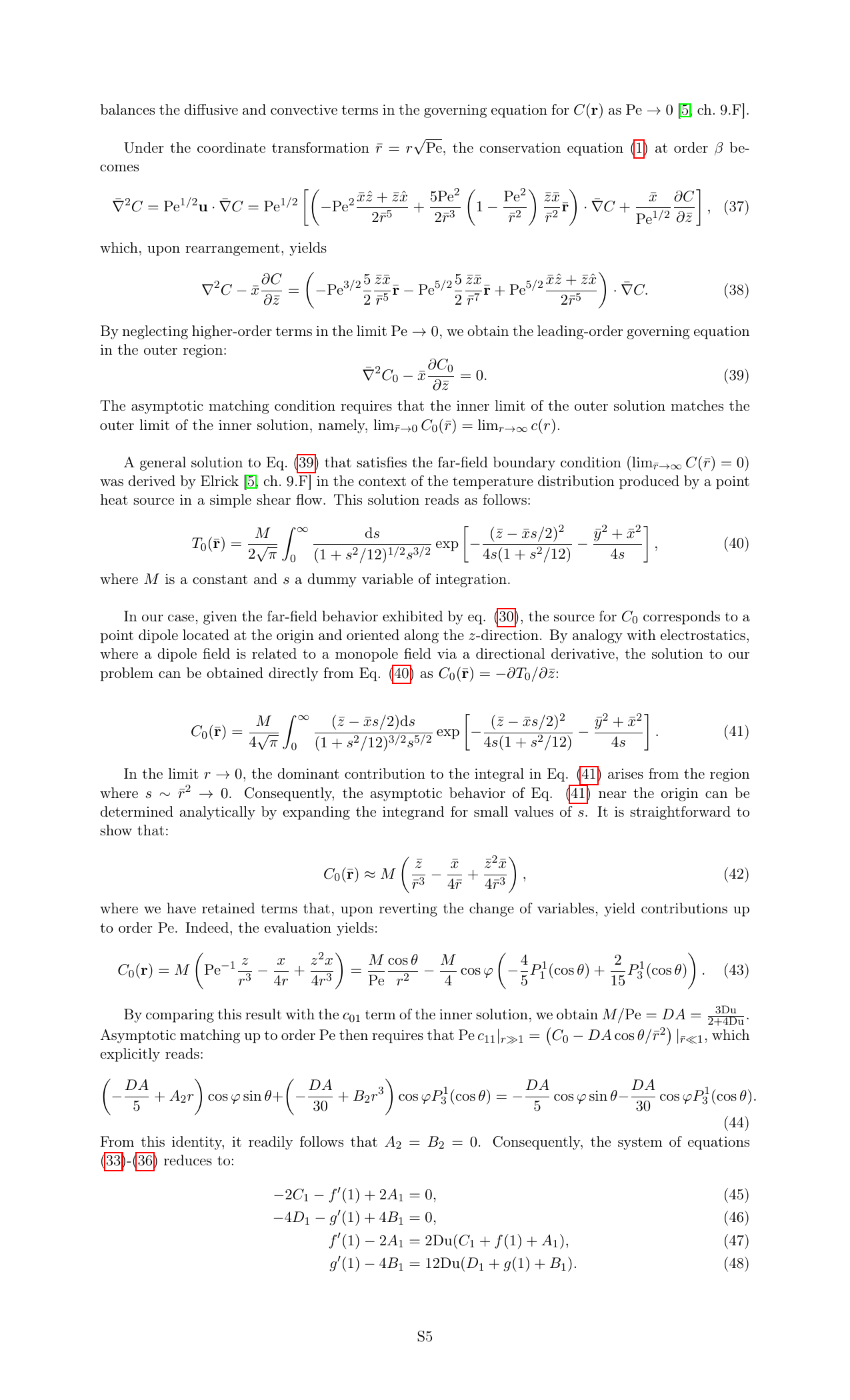}
\end{figure}

\clearpage
\begin{figure}[p]
    \centering
    \includegraphics[width=0.9\textwidth,page=6]{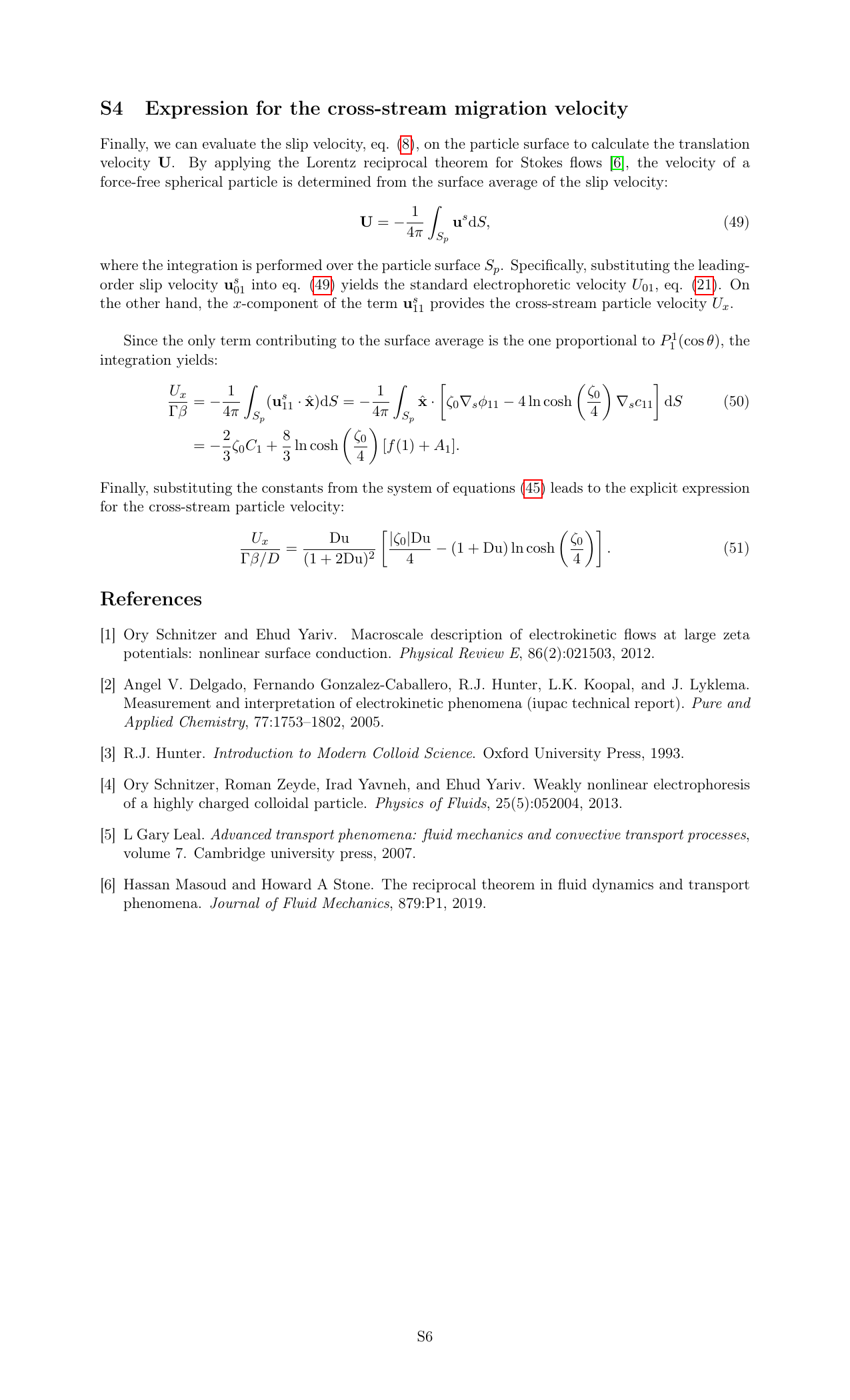}
\end{figure}


\begin{thebibliography}{24}%
\makeatletter
\providecommand \@ifxundefined [1]{%
 \@ifx{#1\undefined}
}%
\providecommand \@ifnum [1]{%
 \ifnum #1\expandafter \@firstoftwo
 \else \expandafter \@secondoftwo
 \fi
}%
\providecommand \@ifx [1]{%
 \ifx #1\expandafter \@firstoftwo
 \else \expandafter \@secondoftwo
 \fi
}%
\providecommand \natexlab [1]{#1}%
\providecommand \enquote  [1]{``#1''}%
\providecommand \bibnamefont  [1]{#1}%
\providecommand \bibfnamefont [1]{#1}%
\providecommand \citenamefont [1]{#1}%
\providecommand \href@noop [0]{\@secondoftwo}%
\providecommand \href [0]{\begingroup \@sanitize@url \@href}%
\providecommand \@href[1]{\@@startlink{#1}\@@href}%
\providecommand \@@href[1]{\endgroup#1\@@endlink}%
\providecommand \@sanitize@url [0]{\catcode `\\12\catcode `\$12\catcode `\&12\catcode `\#12\catcode `\^12\catcode `\_12\catcode `\%12\relax}%
\providecommand \@@startlink[1]{}%
\providecommand \@@endlink[0]{}%
\providecommand \url  [0]{\begingroup\@sanitize@url \@url }%
\providecommand \@url [1]{\endgroup\@href {#1}{\urlprefix }}%
\providecommand \urlprefix  [0]{URL }%
\providecommand \Eprint [0]{\href }%
\providecommand \doibase [0]{https://doi.org/}%
\providecommand \selectlanguage [0]{\@gobble}%
\providecommand \bibinfo  [0]{\@secondoftwo}%
\providecommand \bibfield  [0]{\@secondoftwo}%
\providecommand \translation [1]{[#1]}%
\providecommand \BibitemOpen [0]{}%
\providecommand \bibitemStop [0]{}%
\providecommand \bibitemNoStop [0]{.\EOS\space}%
\providecommand \EOS [0]{\spacefactor3000\relax}%
\providecommand \BibitemShut  [1]{\csname bibitem#1\endcsname}%
\let\auto@bib@innerbib\@empty
%</preamble>
\bibitem [{\citenamefont {Hunter}(1993)}]{hunter}%
  \BibitemOpen
  \bibfield  {author} {\bibinfo {author} {\bibfnamefont {R.}~\bibnamefont {Hunter}},\ }\href@noop {} {\emph {\bibinfo {title} {Introduction to Modern Colloid Science}}}\ (\bibinfo  {publisher} {Oxford University Press},\ \bibinfo {year} {1993})\BibitemShut {NoStop}%
\bibitem [{\citenamefont {von Smoluchowski}(1903)}]{smoluchowski}%
  \BibitemOpen
  \bibfield  {author} {\bibinfo {author} {\bibfnamefont {M.}~\bibnamefont {von Smoluchowski}},\ }\bibfield  {title} {\bibinfo {title} {Contribution {\`a} la th{\'e}orie de l'endosmose {\'e}lectrique et de quelques ph{\'e}nom{\`e}nes corr{\'e}latifs},\ }\href@noop {} {\bibfield  {journal} {\bibinfo  {journal} {Bull. Akad. Sci. Cracovie.}\ }\textbf {\bibinfo {volume} {8}},\ \bibinfo {pages} {182} (\bibinfo {year} {1903})}\BibitemShut {NoStop}%
\bibitem [{\citenamefont {Zhu}\ \emph {et~al.}(2012)\citenamefont {Zhu}, \citenamefont {Lu},\ and\ \citenamefont {Liu}}]{zhu2012}%
  \BibitemOpen
  \bibfield  {author} {\bibinfo {author} {\bibfnamefont {Z.}~\bibnamefont {Zhu}}, \bibinfo {author} {\bibfnamefont {J.~J.}\ \bibnamefont {Lu}},\ and\ \bibinfo {author} {\bibfnamefont {S.}~\bibnamefont {Liu}},\ }\bibfield  {title} {\bibinfo {title} {Protein separation by capillary gel electrophoresis: a review},\ }\href@noop {} {\bibfield  {journal} {\bibinfo  {journal} {Analytica chimica acta}\ }\textbf {\bibinfo {volume} {709}},\ \bibinfo {pages} {21} (\bibinfo {year} {2012})}\BibitemShut {NoStop}%
\bibitem [{\citenamefont {Stone}\ \emph {et~al.}(2004)\citenamefont {Stone}, \citenamefont {Stroock},\ and\ \citenamefont {Ajdari}}]{Stone04}%
  \BibitemOpen
  \bibfield  {author} {\bibinfo {author} {\bibfnamefont {H.}~\bibnamefont {Stone}}, \bibinfo {author} {\bibfnamefont {A.}~\bibnamefont {Stroock}},\ and\ \bibinfo {author} {\bibfnamefont {A.}~\bibnamefont {Ajdari}},\ }\bibfield  {title} {\bibinfo {title} {Engineering flows in small devices: Microfluidics toward a lab-on-a-chip},\ }\href@noop {} {\bibfield  {journal} {\bibinfo  {journal} {Annu. Rev. Fluid Mech.}\ }\textbf {\bibinfo {volume} {36}},\ \bibinfo {pages} {381411} (\bibinfo {year} {2004})}\BibitemShut {NoStop}%
\bibitem [{\citenamefont {Vishwanathan}\ and\ \citenamefont {Juarez}(2023)}]{vishwanathan2023}%
  \BibitemOpen
  \bibfield  {author} {\bibinfo {author} {\bibfnamefont {G.}~\bibnamefont {Vishwanathan}}\ and\ \bibinfo {author} {\bibfnamefont {G.}~\bibnamefont {Juarez}},\ }\bibfield  {title} {\bibinfo {title} {Synchronous oscillatory electro-inertial focusing of microparticles},\ }\href@noop {} {\bibfield  {journal} {\bibinfo  {journal} {Biomicrofluidics}\ }\textbf {\bibinfo {volume} {17}} (\bibinfo {year} {2023})}\BibitemShut {NoStop}%
\bibitem [{\citenamefont {Abdorahimzadeh}\ \emph {et~al.}(2023)\citenamefont {Abdorahimzadeh}, \citenamefont {Pratiwi}, \citenamefont {Vainio}, \citenamefont {Liimatainen},\ and\ \citenamefont {Elbuken}}]{Abdorahimzadeh2023}%
  \BibitemOpen
  \bibfield  {author} {\bibinfo {author} {\bibfnamefont {S.}~\bibnamefont {Abdorahimzadeh}}, \bibinfo {author} {\bibfnamefont {F.~W.}\ \bibnamefont {Pratiwi}}, \bibinfo {author} {\bibfnamefont {S.~J.}\ \bibnamefont {Vainio}}, \bibinfo {author} {\bibfnamefont {H.}~\bibnamefont {Liimatainen}},\ and\ \bibinfo {author} {\bibfnamefont {C.}~\bibnamefont {Elbuken}},\ }\bibfield  {title} {\bibinfo {title} {Interplay of electric field and pressure-driven flow inducing microfluidic particle migration},\ }\href@noop {} {\bibfield  {journal} {\bibinfo  {journal} {Chemical Engineering Science}\ }\textbf {\bibinfo {volume} {276}} (\bibinfo {year} {2023})}\BibitemShut {NoStop}%
\bibitem [{\citenamefont {Abdorahimzadeh}\ \emph {et~al.}(2024)\citenamefont {Abdorahimzadeh}, \citenamefont {Bölükkaya}, \citenamefont {Vainio}, \citenamefont {Liimatainen},\ and\ \citenamefont {Elbuken}}]{Abdorahimzadeh2024}%
  \BibitemOpen
  \bibfield  {author} {\bibinfo {author} {\bibfnamefont {S.}~\bibnamefont {Abdorahimzadeh}}, \bibinfo {author} {\bibfnamefont {Z.}~\bibnamefont {Bölükkaya}}, \bibinfo {author} {\bibfnamefont {S.~J.}\ \bibnamefont {Vainio}}, \bibinfo {author} {\bibfnamefont {H.}~\bibnamefont {Liimatainen}},\ and\ \bibinfo {author} {\bibfnamefont {C.}~\bibnamefont {Elbuken}},\ }\bibfield  {title} {\bibinfo {title} {Anomalous electrohydrodynamic cross-stream particle migration},\ }\href@noop {} {\bibfield  {journal} {\bibinfo  {journal} {Physics of Fluids}\ }\textbf {\bibinfo {volume} {36}} (\bibinfo {year} {2024})}\BibitemShut {NoStop}%
\bibitem [{\citenamefont {Li}\ and\ \citenamefont {Xuan}(2018)}]{li2018}%
  \BibitemOpen
  \bibfield  {author} {\bibinfo {author} {\bibfnamefont {D.}~\bibnamefont {Li}}\ and\ \bibinfo {author} {\bibfnamefont {X.}~\bibnamefont {Xuan}},\ }\bibfield  {title} {\bibinfo {title} {Electrophoretic slip-tuned particle migration in microchannel viscoelastic fluid flows},\ }\href@noop {} {\bibfield  {journal} {\bibinfo  {journal} {Physical Review Fluids}\ }\textbf {\bibinfo {volume} {3}},\ \bibinfo {pages} {074202} (\bibinfo {year} {2018})}\BibitemShut {NoStop}%
\bibitem [{\citenamefont {Di~Carlo}\ \emph {et~al.}(2007)\citenamefont {Di~Carlo}, \citenamefont {Irimia}, \citenamefont {Tompkins},\ and\ \citenamefont {Toner}}]{DiCarlo2007}%
  \BibitemOpen
  \bibfield  {author} {\bibinfo {author} {\bibfnamefont {D.}~\bibnamefont {Di~Carlo}}, \bibinfo {author} {\bibfnamefont {D.}~\bibnamefont {Irimia}}, \bibinfo {author} {\bibfnamefont {R.~G.}\ \bibnamefont {Tompkins}},\ and\ \bibinfo {author} {\bibfnamefont {M.}~\bibnamefont {Toner}},\ }\bibfield  {title} {\bibinfo {title} {Continuous inertial focusing, ordering, and separation of particles in microchannels},\ }\href@noop {} {\bibfield  {journal} {\bibinfo  {journal} {Proceedings of the National Academy of Sciences (PNAS)}\ }\textbf {\bibinfo {volume} {104}},\ \bibinfo {pages} {18892} (\bibinfo {year} {2007})}\BibitemShut {NoStop}%
\bibitem [{\citenamefont {Khair}\ and\ \citenamefont {Kabarowski}(2020)}]{khair2020}%
  \BibitemOpen
  \bibfield  {author} {\bibinfo {author} {\bibfnamefont {A.~S.}\ \bibnamefont {Khair}}\ and\ \bibinfo {author} {\bibfnamefont {J.~K.}\ \bibnamefont {Kabarowski}},\ }\bibfield  {title} {\bibinfo {title} {Migration of an electrophoretic particle in a weakly inertial or viscoelastic shear flow},\ }\href@noop {} {\bibfield  {journal} {\bibinfo  {journal} {Physical Review Fluids}\ }\textbf {\bibinfo {volume} {5}} (\bibinfo {year} {2020})}\BibitemShut {NoStop}%
\bibitem [{\citenamefont {Choudhary}\ \emph {et~al.}(2019)\citenamefont {Choudhary}, \citenamefont {Renganathan},\ and\ \citenamefont {Pushpavanam}}]{Choudhary2019}%
  \BibitemOpen
  \bibfield  {author} {\bibinfo {author} {\bibfnamefont {A.}~\bibnamefont {Choudhary}}, \bibinfo {author} {\bibfnamefont {T.}~\bibnamefont {Renganathan}},\ and\ \bibinfo {author} {\bibfnamefont {S.}~\bibnamefont {Pushpavanam}},\ }\bibfield  {title} {\bibinfo {title} {Inertial migration of an electrophoretic rigid sphere in a two-dimensional poiseuille flow},\ }\href {https://doi.org/10.1017/jfm.2019.479} {\bibfield  {journal} {\bibinfo  {journal} {Journal of Fluid Mechanics}\ }\textbf {\bibinfo {volume} {874}},\ \bibinfo {pages} {856–890} (\bibinfo {year} {2019})}\BibitemShut {NoStop}%
\bibitem [{\citenamefont {Dukhin}\ and\ \citenamefont {Shilov}(1974)}]{dukhinshilovbook}%
  \BibitemOpen
  \bibfield  {author} {\bibinfo {author} {\bibfnamefont {S.}~\bibnamefont {Dukhin}}\ and\ \bibinfo {author} {\bibfnamefont {V.~N.}\ \bibnamefont {Shilov}},\ }\href@noop {} {\emph {\bibinfo {title} {Dielectric phenomena and the double layer in disperse systems and polyelectrolytes}}}\ (\bibinfo  {publisher} {John Wiley and Sons},\ \bibinfo {year} {1974})\BibitemShut {NoStop}%
\bibitem [{\citenamefont {Schnitzer}\ and\ \citenamefont {Yariv}(2012)}]{Schnitzer2012}%
  \BibitemOpen
  \bibfield  {author} {\bibinfo {author} {\bibfnamefont {O.}~\bibnamefont {Schnitzer}}\ and\ \bibinfo {author} {\bibfnamefont {E.}~\bibnamefont {Yariv}},\ }\bibfield  {title} {\bibinfo {title} {Macroscale description of electrokinetic flows at large zeta potentials: nonlinear surface conduction},\ }\href@noop {} {\bibfield  {journal} {\bibinfo  {journal} {Physical Review E}\ }\textbf {\bibinfo {volume} {86}},\ \bibinfo {pages} {021503} (\bibinfo {year} {2012})}\BibitemShut {NoStop}%
\bibitem [{\citenamefont {Schnitzer}\ \emph {et~al.}(2013)\citenamefont {Schnitzer}, \citenamefont {Zeyde}, \citenamefont {Yavneh},\ and\ \citenamefont {Yariv}}]{Schnitzer2013PhysFluids}%
  \BibitemOpen
  \bibfield  {author} {\bibinfo {author} {\bibfnamefont {O.}~\bibnamefont {Schnitzer}}, \bibinfo {author} {\bibfnamefont {R.}~\bibnamefont {Zeyde}}, \bibinfo {author} {\bibfnamefont {I.}~\bibnamefont {Yavneh}},\ and\ \bibinfo {author} {\bibfnamefont {E.}~\bibnamefont {Yariv}},\ }\bibfield  {title} {\bibinfo {title} {Weakly nonlinear electrophoresis of a highly charged colloidal particle},\ }\href@noop {} {\bibfield  {journal} {\bibinfo  {journal} {Physics of Fluids}\ }\textbf {\bibinfo {volume} {25}},\ \bibinfo {pages} {052004} (\bibinfo {year} {2013})}\BibitemShut {NoStop}%
\bibitem [{\citenamefont {Schnitzer}\ and\ \citenamefont {Yariv}(2014)}]{Schnitzer2014PhysFluids}%
  \BibitemOpen
  \bibfield  {author} {\bibinfo {author} {\bibfnamefont {O.}~\bibnamefont {Schnitzer}}\ and\ \bibinfo {author} {\bibfnamefont {E.}~\bibnamefont {Yariv}},\ }\bibfield  {title} {\bibinfo {title} {Nonlinear electrophoresis at arbitrary field strengths: small-dukhin-number analysis},\ }\href@noop {} {\bibfield  {journal} {\bibinfo  {journal} {Physics of Fluids}\ }\textbf {\bibinfo {volume} {26}},\ \bibinfo {pages} {122002} (\bibinfo {year} {2014})}\BibitemShut {NoStop}%
\bibitem [{\citenamefont {Calero}\ \emph {et~al.}(2021)\citenamefont {Calero}, \citenamefont {Fern{\'a}ndez-Mateo}, \citenamefont {Morgan}, \citenamefont {Garc{\'\i}a-S{\'a}nchez},\ and\ \citenamefont {Ramos}}]{Calero2021PRAppl}%
  \BibitemOpen
  \bibfield  {author} {\bibinfo {author} {\bibfnamefont {V.}~\bibnamefont {Calero}}, \bibinfo {author} {\bibfnamefont {R.}~\bibnamefont {Fern{\'a}ndez-Mateo}}, \bibinfo {author} {\bibfnamefont {H.}~\bibnamefont {Morgan}}, \bibinfo {author} {\bibfnamefont {P.}~\bibnamefont {Garc{\'\i}a-S{\'a}nchez}},\ and\ \bibinfo {author} {\bibfnamefont {A.}~\bibnamefont {Ramos}},\ }\bibfield  {title} {\bibinfo {title} {Stationary electro-osmotic flow driven by ac fields around insulators},\ }\href@noop {} {\bibfield  {journal} {\bibinfo  {journal} {Physical Review Applied}\ }\textbf {\bibinfo {volume} {15}},\ \bibinfo {pages} {014047} (\bibinfo {year} {2021})}\BibitemShut {NoStop}%
\bibitem [{\citenamefont {Fern{\'a}ndez-Mateo}\ \emph {et~al.}(2021)\citenamefont {Fern{\'a}ndez-Mateo}, \citenamefont {Garc{\'\i}a-S{\'a}nchez}, \citenamefont {Calero}, \citenamefont {Morgan},\ and\ \citenamefont {Ramos}}]{fernandez2021}%
  \BibitemOpen
  \bibfield  {author} {\bibinfo {author} {\bibfnamefont {R.}~\bibnamefont {Fern{\'a}ndez-Mateo}}, \bibinfo {author} {\bibfnamefont {P.}~\bibnamefont {Garc{\'\i}a-S{\'a}nchez}}, \bibinfo {author} {\bibfnamefont {V.}~\bibnamefont {Calero}}, \bibinfo {author} {\bibfnamefont {H.}~\bibnamefont {Morgan}},\ and\ \bibinfo {author} {\bibfnamefont {A.}~\bibnamefont {Ramos}},\ }\bibfield  {title} {\bibinfo {title} {Stationary electro-osmotic flow driven by ac fields around charged dielectric spheres},\ }\href@noop {} {\bibfield  {journal} {\bibinfo  {journal} {Journal of Fluid Mechanics}\ }\textbf {\bibinfo {volume} {924}},\ \bibinfo {pages} {R2} (\bibinfo {year} {2021})}\BibitemShut {NoStop}%
\bibitem [{\citenamefont {Delgado}\ \emph {et~al.}(2005)\citenamefont {Delgado}, \citenamefont {Gonzalez-Caballero}, \citenamefont {Hunter}, \citenamefont {Koopal},\ and\ \citenamefont {Lyklema}}]{delgado05}%
  \BibitemOpen
  \bibfield  {author} {\bibinfo {author} {\bibfnamefont {A.~V.}\ \bibnamefont {Delgado}}, \bibinfo {author} {\bibfnamefont {F.}~\bibnamefont {Gonzalez-Caballero}}, \bibinfo {author} {\bibfnamefont {R.}~\bibnamefont {Hunter}}, \bibinfo {author} {\bibfnamefont {L.}~\bibnamefont {Koopal}},\ and\ \bibinfo {author} {\bibfnamefont {J.}~\bibnamefont {Lyklema}},\ }\bibfield  {title} {\bibinfo {title} {Measurement and interpretation of electrokinetic phenomena (iupac technical report)},\ }\href@noop {} {\bibfield  {journal} {\bibinfo  {journal} {Pure and Applied Chemistry}\ }\textbf {\bibinfo {volume} {77}},\ \bibinfo {pages} {1753} (\bibinfo {year} {2005})}\BibitemShut {NoStop}%
\bibitem [{\citenamefont {Prieve}\ \emph {et~al.}(1984)\citenamefont {Prieve}, \citenamefont {Anderson}, \citenamefont {Ebel},\ and\ \citenamefont {Lowell}}]{Prieve1984JFM}%
  \BibitemOpen
  \bibfield  {author} {\bibinfo {author} {\bibfnamefont {D.}~\bibnamefont {Prieve}}, \bibinfo {author} {\bibfnamefont {J.}~\bibnamefont {Anderson}}, \bibinfo {author} {\bibfnamefont {J.}~\bibnamefont {Ebel}},\ and\ \bibinfo {author} {\bibfnamefont {M.}~\bibnamefont {Lowell}},\ }\bibfield  {title} {\bibinfo {title} {Motion of a particle generated by chemical gradients. part 2. electrolytes},\ }\href@noop {} {\bibfield  {journal} {\bibinfo  {journal} {Journal of Fluid Mechanics}\ }\textbf {\bibinfo {volume} {148}},\ \bibinfo {pages} {247} (\bibinfo {year} {1984})}\BibitemShut {NoStop}%
\bibitem [{\citenamefont {Yariv}\ \emph {et~al.}(2011)\citenamefont {Yariv}, \citenamefont {Schnitzer},\ and\ \citenamefont {Frankel}}]{yariv2011}%
  \BibitemOpen
  \bibfield  {author} {\bibinfo {author} {\bibfnamefont {E.}~\bibnamefont {Yariv}}, \bibinfo {author} {\bibfnamefont {O.}~\bibnamefont {Schnitzer}},\ and\ \bibinfo {author} {\bibfnamefont {I.}~\bibnamefont {Frankel}},\ }\bibfield  {title} {\bibinfo {title} {Streaming-potential phenomena in the thin-debye-layer limit. part 1. general theory},\ }\href@noop {} {\bibfield  {journal} {\bibinfo  {journal} {Journal of fluid mechanics}\ }\textbf {\bibinfo {volume} {685}},\ \bibinfo {pages} {306} (\bibinfo {year} {2011})}\BibitemShut {NoStop}%
\bibitem [{\citenamefont {Leal}(2007)}]{leal2007}%
  \BibitemOpen
  \bibfield  {author} {\bibinfo {author} {\bibfnamefont {L.~G.}\ \bibnamefont {Leal}},\ }\href@noop {} {\emph {\bibinfo {title} {Advanced transport phenomena: fluid mechanics and convective transport processes}}},\ Vol.~\bibinfo {volume} {7}\ (\bibinfo  {publisher} {Cambridge university press},\ \bibinfo {year} {2007})\BibitemShut {NoStop}%
\bibitem [{\citenamefont {Masoud}\ and\ \citenamefont {Stone}(2019)}]{masoud2019}%
  \BibitemOpen
  \bibfield  {author} {\bibinfo {author} {\bibfnamefont {H.}~\bibnamefont {Masoud}}\ and\ \bibinfo {author} {\bibfnamefont {H.~A.}\ \bibnamefont {Stone}},\ }\bibfield  {title} {\bibinfo {title} {The reciprocal theorem in fluid dynamics and transport phenomena},\ }\href@noop {} {\bibfield  {journal} {\bibinfo  {journal} {Journal of Fluid Mechanics}\ }\textbf {\bibinfo {volume} {879}},\ \bibinfo {pages} {P1} (\bibinfo {year} {2019})}\BibitemShut {NoStop}%
\bibitem [{\citenamefont {Ermolina}\ and\ \citenamefont {Morgan}(2005)}]{ermolina05}%
  \BibitemOpen
  \bibfield  {author} {\bibinfo {author} {\bibfnamefont {I.}~\bibnamefont {Ermolina}}\ and\ \bibinfo {author} {\bibfnamefont {H.}~\bibnamefont {Morgan}},\ }\bibfield  {title} {\bibinfo {title} {The electrokinetic properties of latex particles: comparison of electrophoresis and dielectrophoresis},\ }\href@noop {} {\bibfield  {journal} {\bibinfo  {journal} {Journal of colloid and interface science}\ }\textbf {\bibinfo {volume} {285}},\ \bibinfo {pages} {419} (\bibinfo {year} {2005})}\BibitemShut {NoStop}%
\bibitem [{\citenamefont {Fern{\'a}ndez-Mateo}\ \emph {et~al.}(2022)\citenamefont {Fern{\'a}ndez-Mateo}, \citenamefont {Calero}, \citenamefont {Morgan}, \citenamefont {Garc{\'\i}a-S{\'a}nchez},\ and\ \citenamefont {Ramos}}]{fernandez2022}%
  \BibitemOpen
  \bibfield  {author} {\bibinfo {author} {\bibfnamefont {R.}~\bibnamefont {Fern{\'a}ndez-Mateo}}, \bibinfo {author} {\bibfnamefont {V.}~\bibnamefont {Calero}}, \bibinfo {author} {\bibfnamefont {H.}~\bibnamefont {Morgan}}, \bibinfo {author} {\bibfnamefont {P.}~\bibnamefont {Garc{\'\i}a-S{\'a}nchez}},\ and\ \bibinfo {author} {\bibfnamefont {A.}~\bibnamefont {Ramos}},\ }\bibfield  {title} {\bibinfo {title} {Wall repulsion of charged colloidal particles during electrophoresis in microfluidic channels},\ }\href@noop {} {\bibfield  {journal} {\bibinfo  {journal} {Physical Review letters}\ }\textbf {\bibinfo {volume} {128}},\ \bibinfo {pages} {074501} (\bibinfo {year} {2022})}\BibitemShut {NoStop}%
\end{thebibliography}
\end{document}